# Beyond QWERTY: A Pressure-Based Text Input Approach for XR that Enables a Touch-Typing like Experience

**Fabian Rücker**
Fraunhofer IGD
ORCID: 0000-0003-4071-8642
fabian.ruecker@igd.fraunhofer.de

**Torben Storch**
Fraunhofer IGD
ORCID: 0009-0004-4940-1478
torben.storch@igd.fraunhofer.de

July 28, 2025

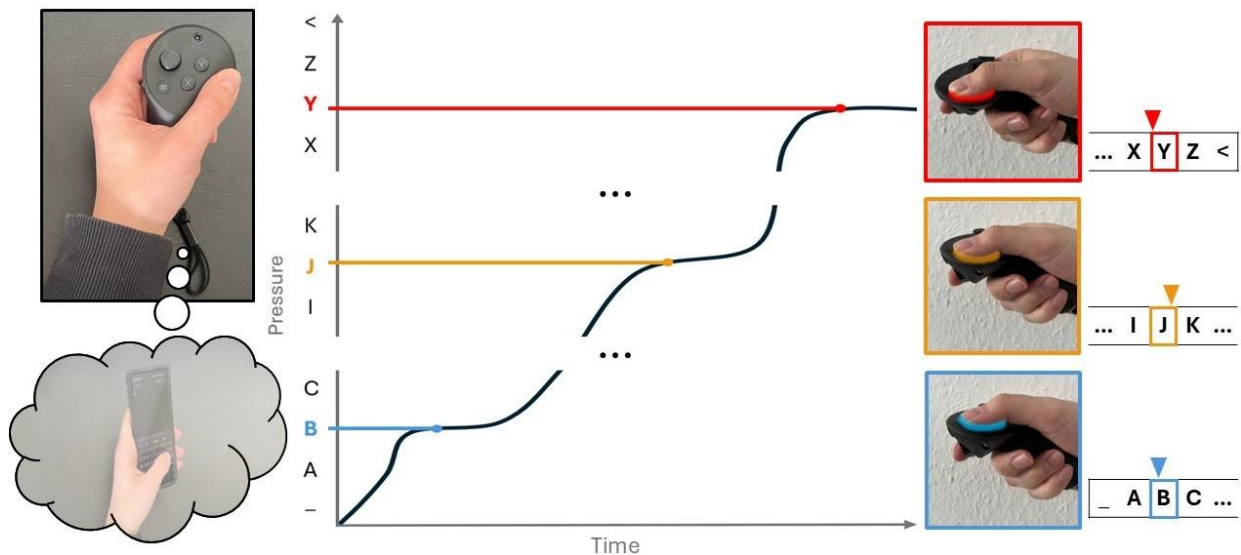

Figure 1: Pressure based text input. Selecting characters by applying varying amounts of pressure to an input device.

## Abstract

Text input in extended reality (XR) applications remains inefficient and tedious. Most solutions are derived from the traditional keyboard layout, yet fail to translate its positive characteristics to the spatial digital realm. This limits the productive use of immersive technologies.

In this work, we analyze physical keyboard input to identify key characteristics that facilitate its comfort, touch typing and high typing speeds. Building on these findings, we propose a novel pressure-based text input modality that transfers these characteristics into immersive space by substituting the two-dimensional QWERTY layout with a linear scale. This design facilitates a touch-typing-like experience, eliminating the need for visual guidance for proficient users.

Our skill-based approach enables typing speeds of over 200 characters per minute. Additionally, it is suitable for discreet use in public spaces and everyday text-input tasks, since the proposed system requires virtually no hand or finger movements and resembles smartphone-based text input in appearance.

# 1 Introduction

Text input poses significant challenges in the field of immersive technologies due to the speed and comfort limitations of established XR input modalities. Although newly introduced XR text input methods have improved typing speeds, they still fall drastically short of the efficiency and comfort offered by traditional keyboards. Consequently, XR text input remains a major inconvenience, making even simple tasks like entering an email address time-consuming and leading many applications to avoid text input altogether.

This challenge becomes increasingly apparent when envisioning the future widespread use of XR devices in everyday scenarios. Established XR input modalities are impractical for usage in public spaces or while on the move. We conclude that spatial digitality necessitates a new input modality, one that fundamentally diverges from the physical keyboard and its familiar QWERTY layout. The two-dimensional nature of QWERTY will always require spatial navigation, causing users to focus more on the virtual input user interface (UI) than on their text output.

**Aim of This Work**   This work aims to identify and outline the fundamental characteristics of keyboard-based text input that facilitate high typing speeds and ensure a comfortable user experience. By analyzing these characteristics, we demonstrate that current XR text-input approaches are inherently incapable of incorporating them successfully. Furthermore, we explain why existing keyboard typing skills do not readily translate to immersive XR applications.

**Analysis of Keyboard Advantages**   A significant portion of traditional keyboard efficiency and comfort stems from extensive practice and the development of muscle memory. Additionally, the spatial design of physical keyboards plays a crucial role in enhancing text input performance. Physical keyboards support touch-typing, enabling users to input text without visual guidance due to haptic cues such as the "homing bars" on the "F" and "J" keys. These haptic cues facilitate continuous recalibration, allowing for non-visual orientation on the two-dimensional layout. Consequently, users can blindly navigate the physical keyboard while focusing solely on the written output, thereby enhancing both typing speed and accuracy.

In contrast, XR text input relies on users continuously referencing the virtual keyboard interface. Without non-visual cues to disclose the cursor's position on the two-dimensional QWERTY layout, a touch-typing-like experience becomes impossible due to accumulating drift. This continuous need to reference the input UI disrupts the typing flow and forces users to focus more on the input rather than the output, significantly reducing overall efficiency.

**Proposed Solution**   To overcome these limitations, we propose an alternative text input modality that preserves the advantages of physical keyboards while adapting to the unique requirements of XR. Although conceptually inspired by Morse code, our system does not rely on fixed "short"/"long" codes for each character. Instead, users select from 28 characters, including the alphabet, space, and backspace, by applying specific pressure levels to the input device. This method allows users to navigate a linear scale of characters, thereby removing the need for spatial navigation that is inherent in QWERTY layouts. By allowing users to select characters through pressure levels, our system facilitates a touch-typing-like experience, as we will explain in 3.

Unlike current XR text input solutions, our system is designed for discreet use in public settings and everyday tasks, closely resembling thumb-based text input on smartphones, which does not require extensive spatial movements either (compare 2). This enhances typing comfort and efficiency within XR contexts. Although the proposed method requires users to learn a new fine motor skill, it can enable proficient users to type non-visually and comfortably with typing speeds exceeding 200 characters per minute with a single hand.

**Rationale for Departing from QWERTY**   Historically speaking, the QWERTY layout has always prevailed: While optimized keyboard layouts like Dvorak offer ergonomics and efficiency improvements, their adoption has been limited due to the high effort required to learn it. This indicates that either the inconvenience of existing input modalities or the advantages of an alternative must be considerably high to justify transitioning.

We believe that the inefficiencies of established XR text input modalities, could provide enough incentive to justify learning a new skill. This becomes increasingly evident when envisioning future everyday use of XR technologies, which presents new requirements to their efficient and productive usage. The currently rather low typing speeds of established raycast based text input modalities, does not comply with these high standards. This work identifies the spatial two-dimensional quality of the QWERTY layout as the bottle-neck, because it will always require a likewise spatially selection of characters. Our pressure-based modality on the other hand enables character selection without pointing at or directly interacting with a virtual UI, which results in a more suitable and efficient text input method for immersive technologies.

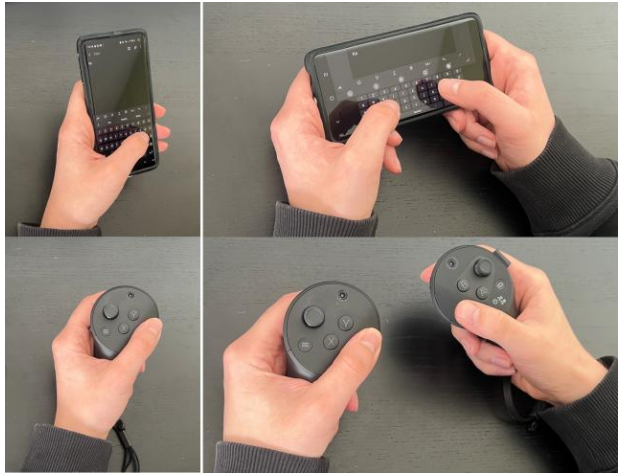

Figure 2: Comparing the smartphone (top) and pressure based (bottom) text input approach, shows the similar appearance. While the thumb needs to navigate on a two-dimensional QWERTY layout on the smartphone, it can comfortably rest in the same location for the proposed input system. Characters are selected by applying different amounts of pressure to the input device.

In summary, the transition from traditional keyboard input to our innovative pressure-based approach represents a significant shift in XR text entry. We are able to retain key-advantages of the physical keyboard at the cost of losing the familiarity of the QWERTY layout. By eliminating the dependence on spatial navigation and enabling non-visual, haptics-based input, our system enhances user comfort and facilitates a touch-typing like experience, that reduces the need to constantly reference the virtual text input UI and allows to focus on the output instead. This paves the way for creating an XR text input solution, that could allow for productive everyday use in the future.

## 2 Related Work

This section begins with an overview of non-immersive text input approaches, such as physical keyboards and smartphones. We then discuss XR text input methods that derive from familiar input modalities, such as the QWERTY layout, followed by alternative solutions like speech-to-text and gesture-based text input.

### 2.1 Overview of Text Input & Communication

The history of keyboard layouts dates back to the development of typewriters, which led to the creation of the QWERTY layout. This layout was designed to prevent key jamming by placing frequently used letters apart, thereby intentionally worsening the layout for typing efficiency. Despite the existence of more ergonomic alternatives, such as the Dvorak layout, these innovations have struggled to gain acceptance, as the discomfort of learning a new layout usually outweighs potential benefits. This illustrates that for users to adopt a new input modality, the inconvenience of existing solutions must be significantly high.

Typing speed on physical QWERTY keyboards has been extensively studied. Dhakal et al. analyzed over 136 million keystrokes from 168,000 volunteers, reporting an average performance of 51.6 words per minute (WPM), with some typists exceeding 120 WPM [1].

### 2.2 Smartphone text input

The advent of mobile devices necessitated new text input modalities, such as multi-tap on early mobile phones, allowing users to select characters by tapping a single key multiple times. This method was later replaced by digital QWERTY keyboards on interactive touchscreens, benefiting from users' familiarity with physical keyboards. While the tactile experience differs, the familiarity of the QWERTY layout has facilitated faster input speeds on smartphones. In 2019, Palin et al. reported that the typing speed gap between physical keyboards and smartphones is closing, with participants averaging at 36.2 WPM and some reaching up to 85 WPM [2]. Notably, this is impressive considering that smartphone text input typically relies on just one or two thumbs, compared to the up to ten fingers used on physical keyboards. The high speeds have been partly attributed to effective auto-correction features.

Earlier studies indicate lower speeds. In 2013 Castellucci and MacKenzie found that novices reached a typing speed of 21.4 WPM on average [3]. Other work reports averages of 4.73 WPM for elderly users (2012, [4]), 15.9 WPM (2010, [5]) and 54 WPM for experienced users (2013, [6]). This evolution underscores significant improvements in smartphone typing speeds, emphasizing the benefits of prolonged practice with an input modality.

Our proposed text input modality relates to smartphone typing, as it utilizes one or two thumbs as well. However, our method differs by requiring a new motor skill that is not vision-based and does not require precise spatial movements. This minimizes reliance on hand-eye coordination. Similar to Morse code, where users transmit characters with one finger, our approach emphasizes tactile input and muscle memory over visual feedback.

## 2.3 QWERTY derived XR text input solutions

By now it has become obvious, that the QWERTY layout is the preferred text input UI for most smartphone users. The same seems to apply for XR based text input, where most applications use a virtual keyboard with QWERTY layout whenever text needs to be entered. Nevertheless, the average XR typing speeds are significantly lower than on smartphones and the physical keyboard.

### 2.3.1 Raycast

Raycast-based text input on QWERTY layouts has emerged as a popular method in XR environments. Users point their tracked VR controllers at a virtual keyboard, highlighting characters to enter. This method typically achieves text input speeds between 15 and 22 WPM. Speicher et al. compared various VR text input methods and found raycast text entry to be the most popular, with an average speed of 15.4 WPM [7].

Boletsis et al. conducted a comparative evaluation of four controller-based VR text-input techniques, including raycasting (16.7 WPM) and drum-like keyboard input (21 WPM) [8]. While their findings suggest that drum input is more efficient, it has yet to achieve broader acceptance.

Additional studies report similar typing speed averages of 17.4 WPM [9], 19.8 WPM [10] and 21.4 WPM [11] for raycast based text input.

Wan et al. introduced a modified raycasting method to enhance alphanumeric and special character input by utilizing different buttons on the physical controllers to access various keyboard layouts [12]. Leng et al. presented the Flower Text Entry method, transferring the QWERTY-layout into a flower-shaped keyboard. While novices achieved 17.65 WPM, experts reached up to 30.80 WPM [13].

Lee and Kim modified standard VR controllers to include additional input buttons for each finger, aiming to transfer touch-typing skills from physical keyboards to VR environments [14]. Nevertheless, their approach still relies in part on raycasting.

### 2.3.2 Hand- & Fingertracking approaches

Kim et al. utilized hand tracking to simulate typing on a smartphone screen, projecting a virtual keyboard above the users' hands, who activate individual keys with their thumbs [15].

Similarly, Richardson et al. aim to transfer the physical keyboard typing experience to XR by using hand- and finger-tracking to estimate intended keystrokes while typing on any flat surface. This approach preserves the touch-typing skills and astonishingly shows parity with typing on a physical keyboard [16].

Kuester et al. introduced KITTY, a non-visual text input method based on the QWERTY layout, where users input characters by touching different fingers with their thumb [17]. While this approach does not mimic physical keyboard typing, it provides a touch-typing-like experience as it does not require visual guidance.

Similarly, DigiTouch by Whitmire et al. enabled non-visual text input using a glove-based device for thumb-to-finger touch interactions, achieving an average typing speed of 16 WPM after 20 minutes of practice [18].

Xu et al. presented TipText, a non-visual text input system where users tap different parts of their first finger to type characters, achieving an average input speed of 11.9 WPM [19].

Dube et al. introduced Shapeshifter, a gesture-based text entry technique using a digital thimble worn on the index finger, which tracks finger position and detects touch and pressure reaching 11 WPM [20]. Dube et al. also explored an ultrasonic haptic feedback system to enhance text entry efficiency for mid-air QWERTY input, finding that this feedback significantly improved typing speed and reduced error rates [21].

Instead of utilizing glove-like devices, it is also possible to use electromyography (EMG) wristbands that are able to read muscle signals. Sivakumar et al. published a large dataset that contains EMG measurements of QWERTY text input on physical keyboards of 108 users [22]. Fu et al. assessed typing speeds of users, who wore an EMG wristband and typed on a T9 keyboard reaching an average typing speed of 15.7 WPM [23]. Meta is currently developing an EMG wristband for their spatial computing platform, which could provide the required accuracy to accomodate for the proposed pressure based text input system.

### 2.3.3 Touchscreen & Touchpad

Utilizing the touchpads of the HTC Vive wands, users can input text by moving a cursor on a split QWERTY keyboard, achieving an average speed of 10.8 WPM [8]. Son et al. improved two-thumb touchpad typing by mimicking smartphone input closely, reaching speeds of up to 20.6 WPM [24].

Boustila et al. demonstrated the use of smartphone touchscreens for VR text input, achieving 12 WPM [11]. The virtual interface mirrored the real touchscreen, allowing fingertip tracking for input.

### 2.4 Non-QWERTY text entry approaches

Not all XR text input paradigms rely on the familiar QWERTY keyboard layout. Twiddler, for example, is a chorded keyboard that allows users to press multiple keys simultaneously with one hand for efficient text input. Lyons et al. found that while Twiddler may be slower initially, it ultimately surpasses multi-tap input methods in speed [25].

Digitap utilizes a wrist-mounted camera to observe fingertaps between the thumb and other fingers, enabling non-visual text input through a multi-tap approach [26].

Smith et al. examined voice based text input. They conducted a comparative study of physical keyboards, smartphones and voice text input, finding that both voice input and physical keyboards were rated highest by participants [27].

Zhu et al. introduced a non-visual text input modality where users draw gestures on a touch-enabled remote control, achieving an average speed of 22 WPM within 10 minutes of practice [28].

## 3 Transferring keyboard advantages to XR

This section provides a comprehensive analysis of text input via physical keyboards, identifying key characteristics that enable comfortable writing at high speeds while maintaining a low error rate. Understanding these aspects is essential, as they will guide our investigation into whether current XR text input approaches effectively incorporate these advantages. By leveraging our findings and building upon prior research, we aim to design a novel input modality that accommodates the identified characteristics requiring us to move away from the familiar QWERTY layout.

### 3.1 Text input analysis

### 3.1.1 Physical Keyboard

Identifying the main advantages of keyboard text input is challenging, as typing on physical keyboards has become second nature for many users. This familiarity can obscure the inherent features that contribute to typing efficiency and accuracy, making it difficult to discern which aspects are superior by design and which are a result of extensive practice. Consequently, comparing new input modalities to traditional keyboards is inherently challenging, as the typing experience differs significantly and requires conscious effort. This discrepancy usually results in newly introduced text input methods lagging significantly behind keyboard efficiency. Consequently, as discussed in 2.2, the average smartphone typing speed has increased significantly throughout the years due to extensive practice.

To analyze the keyboard's advantages, it is essential to observe the thought processes and motor actions involved in typing. Below is a concise overview of the sequential process involved in typing a single character on a keyboard:

1. Conceptualize the desired character.
2. Determine its location on the keyboard.
3. Move the corresponding finger towards the respective key.
4. Activate the key.

Note, that this process runs in parallel for multiple keys with individual finger movements at the same time. When writing a word usually the respective fingers are already moving towards the subsequent characters to allow for faster key-activation succession. It becomes obvious that writing a word or phrase resembles accessing a well-rehearsed pattern rather than selecting individual keys. The process of performing the aforementioned four distinct parts is largely subconscious for experienced users.

While the first step translates easily to XR input methods, the second step (locating characters) relies on the use of the familiar QWERTY layout. The subsequent steps, however, are not retained when using alternative input modalities, as muscle memory developed through keyboard usage does not carry over to an input device like an XR controller.

We have identified three main advantages of typing with a physical keyboard that we want to examine further: the ability to touch-type without looking, the ability to keep both hands in a comfortable position while typing and the ability to write fast and efficiently.

**Touch-typing system**  The physical keyboard supports touch typing, allowing users to type without looking at the keyboard after initially placing their hands on it. Regular non-visual recalibration is facilitated by two haptic cues, known as "homing bars", located on the "f" and "j" keys, which eliminate the need for users to frequently glance at their keyboard.

This enables users to concentrate on their output, focusing solely on the written text and correcting mistakes based on what appears in front of them. In contrast, XR input methods typically require users to focus on the input UI.

The consistency of the calibrated starting position fosters muscle memory, as users learn the movements of individual fingers based on their familiarity with the keyboard layout.

**Comfort**  Typing on a physical keyboard allows both hands to rest comfortably, requiring only the fingers to move to activate different keys. This relaxed position significantly minimizes fatigue compared to XR input, as users do not need to lift their hands for extended periods. However, switching between a keyboard and mouse can be cumbersome and distracting, particularly for tasks that require frequent transitions.

The aforementioned touch-typing capability further enhances comfort. Since there is no need to focus on the input device or even the output, users can gaze away, allowing their minds to engage in creative thought while continuing to type.

**Efficiency**  Physical keyboards allow users to type efficiently with all ten fingers, maximizing speed and fluidity in the writing experience. This enables them to produce text rapidly, enhancing overall productivity. In contrast, XR input modalities usually rely on a maximum of two independent input devices.

However, a large part of typing efficiency can be attributed to the excessive amount of practice that many users have with physical keyboards as the closing text input gap between keyboard and smartphone proves [2].

### 3.1.2   Smartphone text input

Smartphone text input represents a more recent input modality compared to the physical keyboard. Users have adapted to this new modality over the years, becoming increasingly proficient. While studies showed that typing on a smartphone is still generally slower than on a physical keyboard [1, 2], it is sufficient for the tasks typically performed on mobile devices.

One of the advantages of smartphone input is that it retains the users' familiarity with the QWERTY layout, allowing them to leverage their existing key-finding skills. Although the physical keyboard is replaced by a virtual one, the layout remains recognizable, which aids in the transition to this new modality.

Typically, users rely on one or two fingers, most commonly their thumbs, to enter text. This means their potential text input speed is reduced by utilizing only two fingers. The base of their hand remains relaxed, which is comparable to the comfort of typing on a physical keyboard.

When comparing the touch-typing capabilities of both modalities, it is interesting to see, that smartphone typing allows for some kind of a calibration simply by holding the phone in hand, which acts as an anchor. It provides a general awareness of the thumb's position relative to the screen, eliminating the need for constant visual reassurance during typing. However, this form of recalibration is not as reliable as the haptic homing bars on a keyboard. Consequently, blind typing is only partially possible and requires the aid of sophisticated auto-correction algorithms. In return, since the input UI is rather small and in close proximity to the output, only minimal eye movements are required to make sure the written text matches with the intended input.

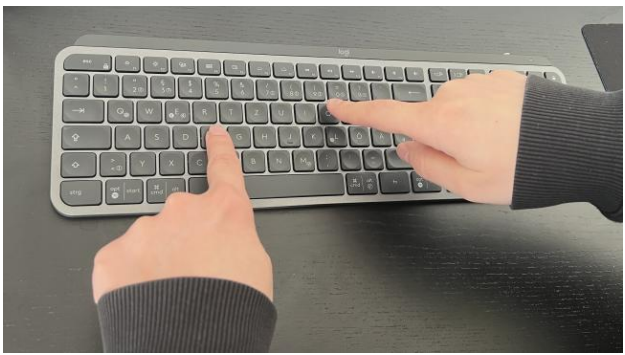

Figure 3: Raycast based XR text input resembles text input with two fingers on a physical keyboard, as it does not allow for a comfortable hand placement and requires frequent focus shifts between the in- and output.

Applying our previous analysis to smartphone text input reveals that step 1 - conceptualization and 2 - character localization remain the same, while steps 3 (moving the finger to the respective key) and 4 (activating the key) differ due to the nature of the input method.

Despite the disruption caused by smartphones, users were generally unwilling to abandon the traditional keyboard layout for an alternative input solution, as it translates sufficiently well to mobile devices for the tasks that need to be performed.

However, we think that in contrast to smartphones, the novelty and challenges of the spatial digital realm introduced by XR technologies justifies the effort of learning a completely new text input modality, because current solutions are tedious. With an effective design, XR text input could be more comfortable and efficient.

In summary, while smartphone text input may not match the speed and efficiency of traditional keyboard typing, it has proven adequate for mobile communication. Comparing early typing speeds on smartphones with current typing speeds shows how practice has improved the performance of this input method over the years [5, 3, 6, 4, 2]. This factor should be considered when conducting user studies with new input modalities.

## 3.2 Translating keyboard skills to XR text input

The primary challenge in translating established typing skills to immersive environments is that the typing experience vastly differs from physical typing. While some dimensions of typing can be transferred to XR, motor skills that are deeply ingrained in muscle memory, such as finger movements, cannot be preserved without a physical keyboard or an input interface that is closely derived from it [16].

Therefore, many XR text input approaches utilize a standard keyboard layout or variations of it to maintain user familiarity. However, we believe that the raycasting approach (see 2.3.1) is not able to achieve similar levels of comfort and efficiency as traditional keyboards or even smartphone text input.

Raycasting requires users to continuously reference the input UI since they need to aim at the individual characters. This requires them to spatially move and rotate their hands to activate each key on the UI keyboard, which makes it impossible to comfortably rest their hands. The continuous need for a direct line of sight between the head-mounted-display (HMD) and the controller to enable reliable tracking further adds to that problem.

Additionally, there is no non-visual re-calibration functionality that users could use to orient themselves on the virtual keyboard without looking. This lack of blind recalibration means that a touch-typing like experience is unattainable, since the raycasting cursor is floating in 3D space without any anchor. The fact that different apps use differently sized virtual keyboards in varying distances adds to that problem. This lack of standardization and non-visual calibration cues prevents the development of muscle memory, since the same movement pattern can lead to different character entries.

Consequently, entering text via raycast in XR resembles and feels like typing on a physical keyboard with only the two forefingers (compare 3). It requires constantly checking the output for errors and does not allow for a comfortable static hand position.

The fact that many developers refrain from text input in VR completely, leads us to our conclusion that the traditional keyboard layout is inadequate for digital realities. This is why attempts to transfer it to immersive contexts fall short of the original. Therefore, we set out to explore new avenues.

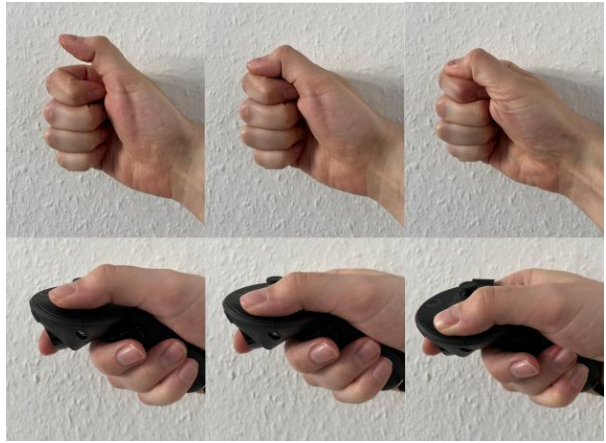

Figure 4: Concept & technical implementation of the proposed input method. From left to right: Rest position, light applied pressure, high applied pressure. The top row depicts the envisioned input method, which could be realized by a digital wristband. The bottom row depicts our technical solution, utilizing a Quest Pro controller that comes with a built-in pressure sensor.

## 4 Pressure based Text Input

In light of our preliminary considerations, our objective is to develop a solution that incorporates the key advantages of traditional keyboard input. As discussed in 3.2, a raycast based typing experience is demanding and cannot facilitate the necessary non-visual recalibration for a touch-typing like experience.

### 4.1 Concept

Our XR text input solution is loosely inspired by Morse code, which allows the transmission of characters with just one finger by sending unique patterns of "long" and "short" signals. However, our system does not discriminate between characters by input patterns but rather by applying different amounts of pressure: characters are represented by specific pressure ranges - the higher the applied pressure, the later the selected character appears in the alphabet. The input interface consists of the linearly arranged alphabet, a live pressure indicator and the currently selected character (see 5). This decouples navigating the UI from the orientation and position of the physical input device, which prevents the drift commonly associated with established XR input methods. Characters are not spatially selected but by reading linear sensor values instead.

We envision a system where users input text solely using their hands, without the need for a hand-held controller. The pressure could be measured by a bracelet like Meta's EMG wristband. By pressing their thumb against the first finger, which provides haptic feedback, users can precisely gauge the applied force. This proprioceptive tactile feedback facilitates precise input, as both the thumb and index finger can detect the pressure.

To ensure reproducible and accurately measurable results, we utilize a Meta Quest Pro controller equipped with sensors to measure the applied pressure (compare 4).

Character input consists of two phases:

**1. Character Selection:** By applying an increasing amount of pressure, the pressure indicator moves through the alphabetic order highlighting each character along the way with a red box. When the desired character is reached, the user reduces the amount of applied pressure, while the desired character stays highlighted.

**2. Character Confirmation:** While the user loosens their grip, the highlighted character remains selected and the pressure indicator moves towards 0 (compare to 6). This allows users to verify their selection before final confirmation. If they wish to correct their input, they can reapply pressure, thus triggering the letter selection process again.

This approach combines steps 3 (moving to a specific character) and 4 (activating it) from our analysis in 3.1 into a single motion. As soon as the applied pressure is sufficient to reach the desired character, it is entered by simply letting go. Since the thumb stays in the same physical location throughout the selection process, the required movement is virtually non-existent.

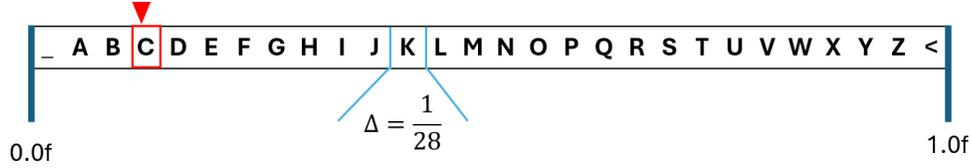

Figure 5: Pressure based input utilizes the linear alphabet for text input. The input interval of $[0, 1]$ is distributed among the selectable characters. For 28 characters, each character has an input interval of size $1/28$. The red triangle indicates the currently applied pressure whereas the red box indicates the currently selected character.

This implies a key difference to established text input solutions: While users are simply able to enter the same key multiple times as soon as their finger hovers above it, this is not the case for our solution, as users need to reapply the same amount of pressure every time. This requires learning a new fine motor skill, which is unfamiliar for first time users and presents an entry barrier. However, this approach allows us to retain the identified advantages of keyboard input.

**Touch-typing**  To effectively achieve touch-typing capabilities, our approach emphasizes the need for a non-visual recalibration system that allows users to reliably navigate the character layout. After each input, users find themselves in the same starting position, similarly to utilizing the homing bars on a physical keyboard. Since the process of selecting a specific character is always the same from start to finish, this approach allows users to develop muscle memory. With practice, they will be able to "feel" each character based on the amount of applied pressure.

**Comfort**  Users can comfortably rest their hands while typing, eliminating the need for spatial movements to select characters. This design mitigates discomfort from holding controllers in tense positions and focusing on virtual UIs, which can lead to fatigue and stress. The hands can rest in any position, and there is no requirement for a direct line of sight between the headset and the controller, as spatial controller tracking is not necessary for reading the pressure sensor's values.

**Efficiency & social acceptance**  The proposed text input system has a high resemblance with smartphone based text input both in appearance and handling. Due to significantly reduced hand movements compared to raycast based text input, our system is suitable for everyday use in public spaces, akin to smartphone typing. Instead of relying on spatial navigation for character selection the thumbs can stay in place. Therefore, the reduced time for spatial thumb movements could make up for the increased input time due to precise pressure application. This needs to be examined in detail.

Overall, the proposed system completely decouples text input from the spatial qualities of the controllers at the price of leaving the familiar QWERTY layout. There is no pointing or spatial interaction with a digital user interface. Instead, users can comfortably apply pressure with their thumbs to select individual characters. Since the thumb always starts and ends in the same position, we can facilitate the development of muscle memory, enabling users to learn how much pressure to apply for each character. This creates a learnable system where users can disregard a spatial physical input device and can instead, once mastered, focus on the output itself.

## 4.2 Technical implementation

This section covers the technical aspects of the proposed text-input approach. It sheds light on some design decisions and how they were implemented. The prototype was developed with Unity 2022.3.16f1, along with the Meta XR All-in-One SDK 68.0.1. The application was tested using Meta Quest Pro controllers with both Meta Quest 3 and Meta Quest Pro headsets. Using the pro controllers was a necessity, since they exclusively offer an input modality called "Thumb Rest Force", which measures how much pressure the thumb applies to the controller. The virtual environment is held simplistic and consists of a graph display showing the pressure level, a text output area and the input alphabet.

### 4.2.1 Basic setup

The proposed text input approach utilizes the linear alphabetical order of characters. This removes the spatial complexity of navigating a two-dimensional array of characters and reduces the problem to a one-dimensional, which can be fully described by a single float value. While we will focus on the thumb rest force input, any input modality that delivers a continuous float value like trigger, shoulder or grip button is suitable for this approach as each delivers a float value between $0.0 f$ and $1.0 f$.

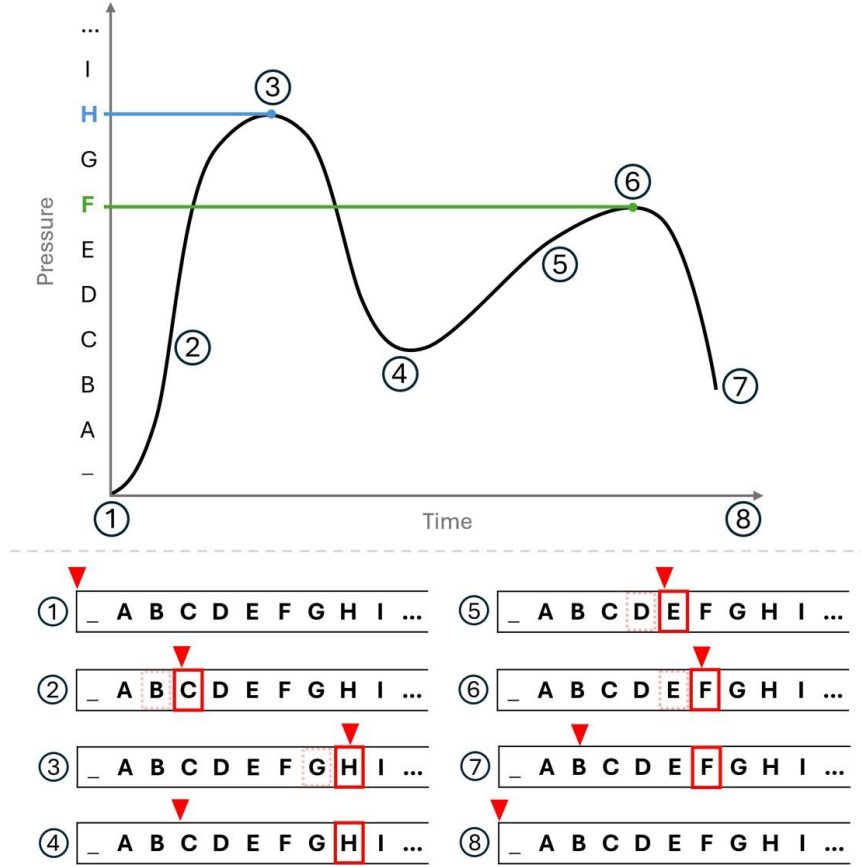

Figure 6: An example of a character selection process showing the pressure graph and the corresponding UI for writing the character "F". The user initially applies too much pressure and overshoots the desired character "F", mistakenly selecting character "H" (1-3), realizes the mistake (4) and reapplies pressure, which reactivates the selection process (5). Lastly the correct letter is selected (6) and entered (7-8).

This range is distributed among the number of typeable characters. In this case we chose to include 28 symbols - 26 letters of the alphabet, space, and backspace. This means that the interval between $0.0 f$ and $1.0 f$ is subdivided in 28 equally sized intervals of length ¹/₂₈.

### 4.2.2 Character selection

The user interface consists of three elements: the alphabet, a triangular indicator that indicates the current amount of applied pressure and a rectangular red box that highlights the currently selected character.

The pressure values are compared on frame-by-frame basis, where the highlighted character is updated, whenever the applied force increased compared to the last frame. If the applied force is decreasing the character highlight is not updated. Characters are highlighted according to their force range within the standardized $[0, 1]$ input interval. When the applied pressure reaches 0, the currently highlighted character is confirmed and entered. 6 depicts an exemplary character selection process that starts with an erroneous selection that is fixed before the wrong character is entered.

### 4.2.3 Input Value Calculation and Character Confirmation

Since displaying the raw sensor data of the thumb rest force input is rather jittery, we chose to incorporate an input buffer that smooths out the raw values and makes character selection more controllable. The buffer averages over three pressure input values at a rate of one per frame and is updated with a first-in-first-out queue. Choosing larger buffer sizes could lead users to timing their input rather than learning the required amount of pressure for each character.

In order to speed up character confirmation we chose to enter the currently highlighted character as soon as the raw input data delivers a value of 0, instead of waiting for the buffer to return to 0. This decision has no downside as a raw value of 0 implies that the user has already released the grip.

Additionally, as 6 shows, it is possible to change the character selection before finally entering it. This can prevent needless character deletions. Comparing this to the physical keyboard and other text input modalities shows, that highlighting a key before entering it might be a meaningful addition to text input in general.

### 4.2.4 Choosing a suitable float input interval

Working with the sensor's raw float values leads to two challenges in the case of pressure based input. Firstly, the sensor activates as soon as users lightly rest their thumbs on the device. Therefore, entering characters requires them to completely let go of the device before the next character can be entered. It is more comfortable to disregard the amount of pressure that is created by resting the thumb, as it allows for continuous character input without completely lifting it. Secondly, reaching high pressure values of $> 0.7$ requires a noticeable amount of force, which makes text input rather demanding over time.

That is why we chose to reduce the relevant input interval to $[0.05, 0.55]$ and remapped it back to $[0, 1]$, where all raw pressure input values $\leq 0.05$ were regarded as $0.0 f$ and all raw values $\geq 0.55$ were regarded as $1.0 f$. This led to a more pleasant user-experience. Nevertheless, by halving the considered input-range, the pressure needs to be applied twice as precisely, effectively halving the pressure range for each character. Therefore, the proposed system provides the capability to fine-tune the experience either toward a more robust or a less demanding character input, which relies on the users' individual preferences.

### 4.2.5 Additional features

**Hold-Delete** Holding down the delete-character, which corresponds to applying maximum force, leads to the continuous removal of characters similar to the holding down the backspace key on keyboards or the smartphone keyboard. This functionality allows for efficient removal of text by holding the delete command, streamlining the editing process.

**Pressure Graph** The pressure graph is updated every frame to provide a real-time visualization of input dynamics. The Y-axis represents the pressure intensity, while the X-axis displays progression of time. When the input value reaches 0, which means the character input is concluded, the graph halts. This visualization helps in monitoring and analyzing pressure input trends in detail.

### 4.2.6 Data Logging

Data logging encompasses various metrics related to input activities. This includes entered character (with exact amount of applied pressure as a float value), per-character input duration, the hand index used for input and the pressure-graph per input. Using these points of data we are able to evaluate each text-input session retrospectively.

## 5   Evaluating general Usability

In contrast to spatial character selection, where users activate keys at specific locations, it is in question whether humans have sufficient fine motor skills to robustly apply different amounts of pressure. Additionally, applying and releasing pressure might be a time-consuming task rendering pressure application for character selection infeasible.

That is why we devised two preliminary experiments, which examine the general usefulness of the proposed system - typing accuracy and speed. Together, they constitute a theoretical lower bound for maximum achievable typing speeds and an upper bound for minimum error rate. These tests ensure that a deliberate use of the proposed text input system is humanly possible. Since the experiments were only performed by one of the authors to determine the aforementioned boundaries, they are not generalizable results for average typing performances of (novice) users.

Both experiments were conducted with only one hand-held controller, which is comparable to typing on the smartphone with one thumb.

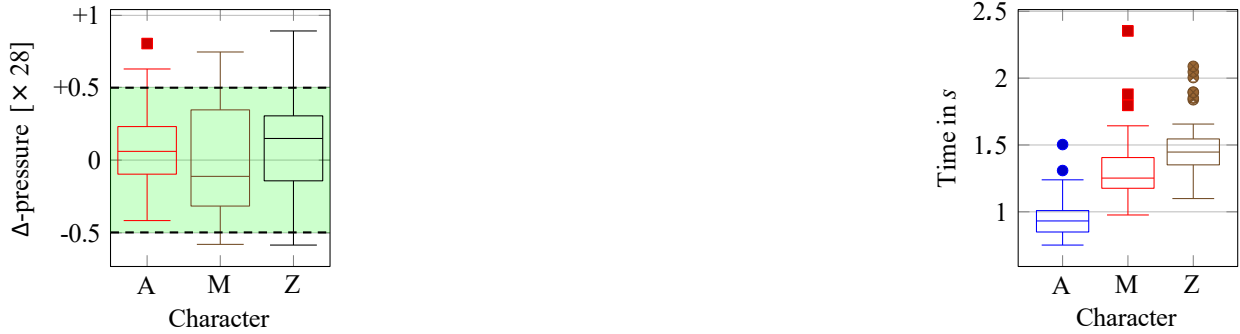

Figure 7: **Accuracy experiment.** This figure shows the results for diligently entering characters. The left hand side shows a box plot of the normalized measured input pressures, where the green area shows the range of acceptable input pressures for each character. Applying a higher or lower pressure leads to a misentry. The right hand side shows a box plot of input times per character in seconds.

## 5.1 Typing Accuracy

To verify the theoretical robustness of our text input system, we decided to choose three characters and carefully entering them for 60 seconds each. The three chosen characters A, M and Z represent the beginning middle and end of the alphabet, which means that each requires a different amount of applied pressure to be entered. The timer started after a character was successfully entered and ended after the last successful entry within a minute. The same conditions were considered for per character timings.

While trying to enter each character correctly, we measured both the input time and the applied pressure per character for this experiment. 7 shows its results.

It becomes evident that characters appearing later in the alphabet take longer to enter on average. This can be attributed to the increased pressure required to move the cursor further. For example, the median input time for the character "A" was $0.93s$, for "M" $1.26s$, and for "Z" $1.45s$. These times correspond to typing speeds of $64.5$, $47.6$, and $41.4$ characters per minute, respectively.

Evaluating the applied input pressure per character shows that it was in general possible to enter the desired characters. The green area in 7 highlights the range of permissible input pressures for the characters. Since each characters input interval is of the same size of $1/28$, we normalized the measured pressure values by subtracting the optimal pressure value for each measurement. This allows for easier comparison between different characters. Most inputs landed within the acceptable interval. Nevertheless, sometimes a neighboring character was erroneously selected an entered. This led to an error rate of $4.76\%$ for "A", $8.7\%$ for "M" and $15\%$ for "Z".

## 5.2 Typing Speed

The second experiment aims to determine a theoretical lower bound for maximum typing speeds achievable with the proposed text input system. We repeated the experiment from before, but focused on entering characters as fast as possible, without considering the correctness of each input.

8 shows the obtained results. As expected the measured pressures scatter significantly more. Nevertheless, 50% of inputs had an inaccuracy of only one character, which means that for all tested characters the selection was either correct or a neighboring character. Still the scattering remains very noticeable with occasionally entering characters that are more than seven positions away from the intended one.

Most input times were below $0.31s$ per character. Many outliers can be attributed to the fact that a character entry was only considered concluded when the pressure value had returned to $0.0f$. Sometimes the pressure was prematurely reapplied before the sensor had a reading of 0, which led to higher input times. Similarly to the precision test, the input time per character increases for characters that occur later in the alphabet - slightly, but continuously. The median input times were $0.24s$ for "A", $0.25s$ for "M" and $0.26s$ for "Z". This corresponds to a theoretical typing speed of $250$, $240$ and $230.8$ characters per minute, respectively.

The error rate is with $63.8\%$ for "A", $81.9\%$ for "M" and $64.2\%$ for "Z" expectably high.

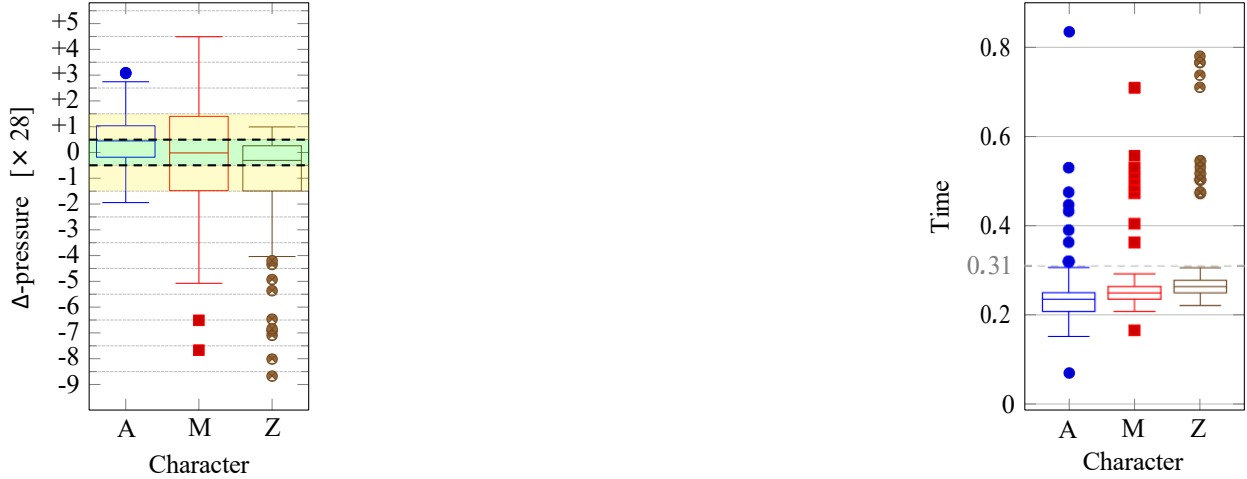

Figure 8: **Speed experiment.** This figure shows the results for entering characters as fast as possible while disregarding accuracy. The left hand side shows a box plot of the normalized measured input pressures, where the green area shows the pressure range for the intended character and the yellow area shows the acceptable pressure ranges for its immediate neighbors. The right hand side shows a box plot of input times per character in seconds.

### 5.3 Additional remarks

The evaluated preliminary experiments verify that the proposed system is willingly usable and assess which typing speeds could theoretically be achieved. However, they do not provide generalizable statistical evidence.

Note, that the experiments would be less meaningful for traditional keyboard based text input, as selecting the same character multiple times means only moving the finger (or cursor) to the respective key once and then activating it again and again. For pressure based text input however, this task is not trivial since users need to re-navigate to the intended character every time. By repeatedly selecting the same character we control the experiment for familiarity with the character-layout. Therefore, it should have no impact whether users are more familiar with the linear alphabet or the two-dimensional QWERTY layout.

This means the experiment simulates typing on a known keyboard layout, where the challenge is reduced to spatially entering characters as fast as possible (e.g. moving fingers to keys). Asking users to type different characters instead of the same one, introduces the additional complexity of finding each character on the layout first, which could distort the results.

## 6 Discussion

We conducted two stress tests on the proposed input modality to estimate its efficiency thresholds, concluding that it merits further examination. While the error rate of up to 15% for careful text input is too high, the accuracy experiment demonstrates that it is indeed possible for users to differentiate between characters by applying varying amounts of pressure. We anticipate that the error rate will decrease as users develop the necessary fine motor skills through practice.

To enhance the user experience, we could consider increasing the pressure range for each character, which would tolerate more inaccuracies during pressure application. Our results indicate that expanding the pressure range by 50% could reduce the total number of errors to zero in the accuracy test, suggesting that the actual pressure inputs were close to the acceptable range.

The speed test serves as an indicator of whether the proposed system can be utilized for rapid text input. Despite the high error rates of up to 81%, we achieved input speeds of less than $0.31$ seconds per character, translating to a typing speed of approximately 193 characters per minute. Additionally, 8 shows that adjusting the pressure range could significantly decrease the error rate. The majority of inputs fell within the range of the intended character and its immediate neighbors. Notably, the distribution for the character "M" is nearly symmetric, with a median value close to zero.

Ultimately, the effectiveness of our system hinges on carefully calibrating the considered pressure range to strike an optimal balance between input reliability and user comfort. Our goal is to ensure robust character selection without imposing too much physical effort on users. One potential improvement involves splitting the alphabet between the left and right hand, similar to the input strategy commonly used in two-thumb smartphone typing, which could effectively double the pressure intervals and reduce accuracy requirements by half.

# 7 Future work

The proposed text input system has not yet been evaluated in a user study. The aim of this work is to establish that the proposed system is motorically viable as a text input solution. The identified lower threshold for maximum typing speed and upper threshold for minimum error rate suggest that the conceived input modality warrants closer examination in future studies.

The planned user study will address questions regarding how quickly users can learn the modality, how fast novice users can type, and how this method compares to established raycast-based text input methods. To enhance the input system's resilience to minor pressure variations, we plan to broaden the range of pressures considered, which should improve overall accuracy. Additionally, we will incorporate autocorrection to address spelling errors, further reducing the required typing precision, akin to smartphone text input methods where users achieve higher typing speeds by increasingly relying on autocorrection algorithms. Given the linear character layout of our proposed system, there are fewer neighboring keys compared to a two-dimensional QWERTY layout, potentially making error detection more straightforward and reliable. Dynamically adjusting the pressure interval sizes for each character based on its probability could further increase accuracy.

We are also interested in identifying optimal pressure thresholds by evaluating the pressure values of resting fingers to determine the minimum pressure required for character selection. Moreover, we aim to ascertain how much pressure users are willing to apply before the system becomes overly demanding, which will help define the relevant float interval of raw data, ensuring an optimal user experience.

Lastly, utilizing novel wristband devices could render the use of a physical handheld controller obsolete. This would make the text input even more proprioceptive, as the thumb presses directly against the forefinger, potentially allowing for more precise pressure application and facilitating accelerated training of muscle memory.

In conclusion, our work sets the stage for future exploration into pressure-based text input systems, which may redefine text entry in XR environments and enhance user interaction with immersive technologies.

# 8 Conclusion

In this work, we presented a novel pressure-based text input system that effectively extends the key advantages of physical keyboards to XR environments by eliminating the need for spatial movement during character selection. This innovation allows users to disregard the physical input device's orientation and position, thereby removing the necessity for a direct line of sight between the HMD and the controllers, which enhances typing comfort. Instead of spatial navigation, characters are selected by applying varying amounts of pressure to the input device. The consistent procedure of selecting characters mirrors the functionality of physical keyboard homing bars, thereby preventing the drift commonly associated with other immersive text input solutions.

The ability to learn specific pressure levels for different characters facilitates the development of muscle memory, enabling non-visual, haptics-based text input. This enables a touch-typing-like experience in XR, making text input more pleasant and allowing experienced users to focus on the written output rather than the input UI. Additionally, the proposed text input modality closely resembles smartphone-based text input and is aesthetically minimalist, making it suitable for everyday use and socially acceptable in public settings.

However, these advantages come with certain challenges. The proposed text input system is skill-based and presents a relatively high entry barrier, requiring users to acquire a new fine motor skill. Similar to chording keyboards, which offer accelerated text input but come with a steep learning curve, our system's initial difficulty might deter potential users. Selecting and entering characters by varying pressure levels is inherently less straightforward than aiming at a character and confirming it with a button press.

Previous innovations in text input, such as optimized keyboard layouts like Dvorak, have not achieved widespread acceptance. This lack of adoption can be attributed not only to the significant effort required to learn a different modality but also to sufficient user satisfaction with existing input methods, rendering the additional effort unjustified. In contrast, text input in XR remains cumbersome, which suggests that the novel spatial digitality and the anticipated demand for efficient XR text input in everyday life could justify the effort of learning a new input modality - even if it necessitates moving away from the familiar QWERTY layout.